\documentclass[12pt]{article}

\def \sect #1 {\setcounter{equation} 0\section{#1}}
\def \be  {\begin{equation}}
\def \ee  {\end{equation}}
\def \ba  {\begin{eqnarray}}
\def \ea  {\end{eqnarray}}
\def \baa {\begin{eqnarray*}}
\def \eaa {\end{eqnarray*}}

\def \as {\relax\ifmmode\alpha_s\else{$\alpha_s${ }}\fi}

\def \LQCD {\Lambda_{\mbox{\tiny QCD}}}
\def \SDG {\mbox{\tiny SDG}}
\def \DGE {\mbox{\tiny DGE}}
\def \eff {\mbox{\tiny eff}}

\def\eq#1{Eq.~(\ref{#1})}
\newcommand{\secn}[1]{Section~\ref{#1}}

\begin{document}

\begin{titlepage}

\rightline{DFTT 15/04}
\rightline{\hfill July 2004}

\vskip 2.5cm

\centerline{\Large \bf Scaling of Power Corrections for Angularities} \vspace*{2mm}
\centerline{\Large \bf from Dressed Gluon Exponentiation} 
 
\vskip 2cm

\centerline{\bf Carola F. Berger$^a$\footnote{e-mail: 
  {\tt carola.berger@to.infn.it}} and Lorenzo Magnea$^b$\footnote{e-mail: 
  {\tt magnea@to.infn.it}}}
\vspace{3mm}
\centerline{\sl $^a$I.N.F.N., Sezione di Torino}
\centerline{\sl Via P.Giuria 1, I--10125 Torino, Italy}
\vspace{3mm}
\centerline{\sl $^b$Dipartimento di Fisica Teorica, Universit\`a di Torino}
\centerline{\sl and I.N.F.N., Sezione di Torino}
\centerline{\sl Via P.Giuria 1, I--10125 Torino, Italy}

\vskip 1.8cm
 
\begin{abstract}

We study power corrections to a recently introduced family of event
shapes, the class of angularities, within the formalism of dressed
gluon exponentiation (DGE). We find that the universal scaling rule
for the leading power corrections deduced from resummation also holds
when taking renormalon enhancements into account. The scaling is due
to boost invariance of eikonal dynamics in the two-jet limit, which we
recover in the context of DGE.  Furthermore, dressed gluon
exponentiation provides an ansatz for non-leading power corrections
that violate the scaling. These non-leading corrections are further
suppressed by non-integer powers of the hard scale.

\end{abstract}

\end{titlepage}

\section{Introduction}
\label{intro}

Event shapes \cite{Farhi:1977sg,weights,broad1} are
infrared-safe generalizations of jet cross sections, and they are
among the best suited observables to test our understanding of
QCD. Event shapes distributions, in particular, describe the behavior
of colored radiation in the final state of hard collisions in all of
phase space. Thus, although infrared safe, they are sensitive to
emission at all scales, and provide a unique tool to probe the
interface between perturbative and non-perturbative QCD.

A consequence of the sensitivity of event shapes to long-distance effects is
that perturbative calculations are far from
straightforward even at high energies, where the strong coupling is sufficiently
small. Specifically, in the narrow-jet limit radiation is
dominated by secondary partons which are either soft or collinear to the
primary quarks emitted at the hard scattering. As a
consequence, fixed order computations receive large logarithmic
corrections, which need to be resummed to all orders to obtain
reliable quantitative predictions
\cite{GS,CTTW1,CTTW2,Dokshitzer:1998kz,Dasgupta:2003iq}.

In addition to these large perturbative corrections, there are
corrections due to confinement that cannot be treated within the
perturbative expansion, and are suppressed by powers of the hard
scale.  Their main effect is to widen the distribution of radiation in
the final state, shifting the peak away from the narrow jet
limit. Although long-distance effects are intrinsically
nonperturbative, the study of resummed perturbation theory can give
valuable information about the size of their contributions to the
cross section, as we will illustrate below.

In the following we will consider event shapes in $e^+e^-$
annihilation. Power corrections to these observables have been the
focus of intense research during the past several years (for recent
reviews see for example \cite{Dasgupta:2003iq,Magrev}). It turns out
that for mean values of such event shapes the effects of power
corrections can be efficiently summarized in terms of a single
parameter, essentially a shift of the perturbative distribution by an
amount proportional to an integer power of $1/Q$, where $Q$ is the
overall center-of-mass energy
\cite{irren,irrdiff,dispers,Gardi:2000yh}. It was found that
these additive corrections are to a certain degree universal. This
universality property can be used phenomenologically for precise
measurements of the strong coupling from event shapes
\cite{alphasmean}.

The case of differential distributions, however, is more
complicated. Distributions probe scales that are even smaller than
those which dominate mean values, thus subleading power corrections
also need to be taken into account. For a large class of such
corrections this can be done by introducing a nonperturbative shape
function \cite{KorSt}. Resummed perturbation theory, which
displays sensitivity to soft emission at power accuracy, imposes
constraints on models for these shape functions. Phenomenologically, the
models are quite successful in describing the data, again with only a
small set of nonperturbative parameters. Universality, however, is
generically lost at the level of subleading power corrections, although
certain classes of event shapes turn out to have closely related
nonperturbative behavior \cite{shapepar,Gardi:2001ny,Gardi:2002bg,GarMag}.

A class of event shapes of this type depending on a real parameter $a$ was introduced in
\cite{BKS1}. These observables provide an interpolation between the thrust \cite{Farhi:1977sg},
corresponding to $a = 0$, and the jet broadening \cite{broad1}, $a =
1$. In this paper, we will refer to event shapes in this class as
\emph{angularities}. As shown in \cite{BKS1,GSParis,CPhD,BS1}, these event shapes can be
studied analytically, as functions of $a$, both at the perturbative
level, performing an all-order resummation of Sudakov logarithms, and
at the level of long-distance effects. Using factorization arguments,
it was found that leading power corrections to all the observables in
the class of angularities are described by the same shape function, up to an overall
scaling factor \cite{BS1}. This remarkable property is closely related
to the boost invariance of the soft radiation emitted in the two-jet
limit.

 We will study power corrections
to the distributions of angularities by using renormalon methods. Specifically, we
will use Dressed Gluon Exponentiation (DGE) \cite{Gardi:2001di} to
build a model of the shape function. We will recover the scaling
property discovered in \cite{BS1}, and we will study the pattern of
violations of scaling due to the contributions of collinear radiation.

Dressed gluon exponentiation combines renormalon calculus with Sudakov
resummation. On the one hand, taking into account single (dressed)
gluon emission in the large $\beta_0$-limit, it identifies the leading
source of factorially divergent behavior, characteristic of the
asymptotic nature of the perturbative expansion. This is then upgraded
to multi-gluon emission in the Sudakov limit by exponentiation of the
single gluon result, employing as usual a Laplace transform to enforce
momentum conservation.  In the conventional resummed expressions, the
asymptotic behavior manifests itself as a singularity in the
perturbative running coupling at small scales, which introduces an
ambiguity in the form of power corrections. The DGE formalism provides
a definite prescription to deal with these ambiguities, and thus
strongly constrains the form of nonperturbative shape functions.
Dressed gluon exponentiation has been previously applied to several
other event shapes, including the thrust and the C-parameter
\cite{Gardi:2001ny,Gardi:2002bg,GarMag,Gardi:1999dq}, as well as to
processes involving heavy quarks \cite{heavy}, and inclusive hadronic
cross sections such as DIS and Drell-Yan~\cite{Gardi:1999dq,dis}. Here
we will apply it to study power corrections to angularities.

An interesting aspect of the study of angularities with this method is
the fact that the scaling rule found in Ref.~\cite{BS1} was shown to
be closely related to the boost invariance properties of the eikonal
cross section describing soft emission in the two-jet limit. This
property is not at all apparent in DGE, where in practice the single
gluon cross section is computed with a ``massive'' gluon, along the
lines of the dispersive approach. Since the introduction of the gluon
virtuality breaks boost invariance, and since in addition is necessary
to account for subleading logarithms in the large $n_f$ limit, it is
not a priori clear whether the scaling will survive. Here we will show
how the effects of boost invariance are recovered in the Sudakov
region. In the two-jet limit, only the logarithmically divergent
bremsstrahlung spectrum contributes to the dressed gluon cross
section, and the gluon mass acts as an effective cutoff, with
precisely the weight required to reconstruct an unweighted rapidity
integral.

We start by briefly reviewing the definition of the class of
angularities, and the scaling rule for nonperturbative corrections
found by analyzing the resummed expression. In Secs. \ref{DGEsec} and \ref{papo} 
we will then construct a model of the shape function for
angularities by means of DGE. Concentrating on soft radiation, we will
recover the scaling of leading power corrections, and study how it
arises in the context of a massive gluon calculation. Finally, we will
observe that DGE suggests an intricate, non-universal pattern of
subleading power corrections arising from collinear radiation. All
such corrections are however suppressed by non-integer powers of the
hard scale, with a degree of suppression growing as the parameter $a$
becomes large and negative, where the event shape becomes correspondingly
more inclusive. Technical details are given in the
Appendix.

Although detailed phenomenological studies will have to be deferred to
future work, when and if experimental data become available, we
emphasize that such studies would be of considerable interest, and
should be quite easy to perform for experimental collaborations.
The scaling rule found in \cite{BS1} and recovered here is in fact a
clean and significant test of the behavior of soft radiation in QCD,
and of the theoretical models employed in recent years to study power
suppressed effects.

\section{The class of angularities} 
\label{Sec2}

We consider an $e^+ e^-$ annihilation event with center-of-mass energy
$Q$, generating a final state $N$, which we will take for now to
consist of massless particles. The angularity, with weight $a$, of the
state $N$, is defined as~\cite{BKS1}
\be
  \tau_a (N) = \frac{1}{Q}\sum_{i \in N} p_{i \perp} \; e^{-|\eta_i| 
  (1 - a)} \, = \frac{1}{Q} \sum_{i \in N} \omega_i \left( \sin 
  \theta_i \right)^a \left( 1 - \left| \cos \theta_i \right| 
  \right)^{1 - a}~,
\label{barfdef}
\ee
where $p_{i \perp}$ is the transverse momentum of particle $i$
relative to the thrust axis, $\eta_i$ is the corresponding
pseudorapidity, $\eta_i = \ln \cot \left( \theta_i/2 \right)$, with
$\theta_i$ the angle with respect to the thrust axis, and $\omega_i$
is the energy of particle $i$.  The two definitions in \eq{barfdef}
are equivalent for massless particles.

The angularity distribution is defined as usual by summing over all
final states, each weighed by its probability, according to
\be
  \frac{d \sigma}{d \tau_a} = {1\over 2 Q^2} \sum_N \;
  |M(N)|^2 \, \delta(\tau_a - \tau_a(N))~,
\label{eventdef}
\ee
where $M(N)$ is the amplitude for the production of final state $N$.

The thrust axis can be defined as the axis with respect to which
\eq{barfdef} is minimized at $a = 0$.  The parameter $a$ is adjustable
in the range $- \infty < a < 2$, with the upper limit set by infrared
safety. Angularity with $a = 0$ is essentially $1 - T$, with $T$ the
thrust \cite{Farhi:1977sg}, while angularity with $a = 1$ corresponds to the jet
broadening \cite{broad1}.  As discussed in~\cite{BKS1}, for $a \geq 1$ recoil
effects become important, so that the resummation of Sudakov
logarithms must be modified, and consequently the pattern of power
corrections changes, as pointed out for the broadening ($a = 1$) 
in \cite{Dokshitzer:1998kz}. In
this paper we will be mostly concerned with the case $a \leq 0$, 
where power corrections are under good control.

In the two-jet limit, $\tau_a \rightarrow 0$, the distribution in 
\eq{eventdef} has large perturbative corrections containing powers of
$\ln \left( \tau_a \right)$, which have been resummed to all logarithmic 
orders, at leading power and for $a < 1$, in \cite{BKS1}. As usual, the
resummation is performed at the level of the Laplace transform of the 
distribution,
\be
  \tilde{\sigma} \left(\nu, a \right) = \int_0^\infty d \tau_a \,
  {\rm e}^{- \nu \tau_a} \, \frac{d \sigma}{d \tau_a}~.
\label{lapla}
\ee
At next-to-leading logarithmic (NLL) level, 
the resummation displays a nontrivial dependence on $a$,
\ba
  \frac{1}{\sigma_{\rm tot}} \, \tilde{\sigma}
  \left(\nu, a \right) & = & 
  \exp \Bigg\{ 2 \, \int\limits_0^1 \frac{d u}{u} \, \Bigg[ \,
  \int\limits_{u^2 Q^2}^{u Q^2} \frac{d p_\perp^2}{p_\perp^2} \,
  A \left(\as (p_\perp) \right)
  \left( {\rm e}^{- u^{1-a} \nu \left(p_\perp/Q\right)^{a} }-1 \right)
  \nonumber \\
  & & \qquad \qquad \quad \quad
  + \,\, \frac{1}{2} \, B\left(\as(\sqrt{u} Q)\right) \left( {\rm e}^{-u
  \left(\nu/2\right)^{2/(2-a)} } -1 \right)
  \Bigg] \Bigg\}~,
\label{thrustcomp}
\ea
where $A(\as)$ and $B(\as)$ are the well-known anomalous dimensions
acting as kernels of Sudakov exponentiation. The intricate $a$
dependence of \eq{thrustcomp} simplifies at the level of leading
logarithms, where one can easily invert the Laplace transform to find
\be
  \frac{1}{\sigma_{\rm tot}} \frac{d \sigma}{d \tau_a} = - \, 
  \frac{2}{1 - a/2} \, \frac{\as}{\pi} \, C_F \, 
  \frac{\ln(\tau_a)}{\tau_a} \,
  \exp \left[- \, \frac{1}{1 - a/2} \,
  \frac{\as}{\pi} \, C_F \, \ln^2 \left( \tau_a \right) \right]~,
\label{LL} 
\ee
which displays a simple scaling with $1 - a/2$. This scaling of the
perturbatively resummed cross section is however only approximate, and
breaks down at NLL level, as can be seen in the explicit expressions
given in the Appendix of Ref.~\cite{BS1}.

Remarkably, the approximate scaling of the perturbative contribution
with $1 - a/2$ is replaced at the level of leading power corrections
by an exact scaling with $1 - a$. To see this, one notes that the
perturbative expression for the cross section given in \eq{thrustcomp}
is ambiguous, due to the fact that the scale of the running coupling
can vanish. As a consequence, at values of $\tau_a \sim \LQCD/Q$
nonperturbative corrections must become dominant, and the perturbative
expression needs to be supplemented by nonperturbative input to give a
well-defined result. The structure of this nonperturbative correction
can be deduced, following \cite{KorSt,BS1}, by introducing
an infrared factorization scale $\kappa$ to cut off the transverse
momentum integration in \eq{thrustcomp}. The leading nonperturbative
contribution arising from small transverse momenta can be evaluated by
performing the integral over the Laplace variable $u$, keeping only
terms scaling as powers of $\nu/Q$, while discarding terms suppressed
by higher powers of the hard scale. The result
can finally be written as a convolution of a perturbative contribution
and a nonperturbative shape function, which in moment space is just a
product,
\ba 
  \tilde{\sigma} \left(\nu, a\right) & = & \tilde{\sigma}_{\rm PT} 
  \left(\nu, \kappa, a\right) \, \tilde{f}_{a, {\rm NP}}
  \left(\frac{\nu}{Q},\kappa\right) \left[ 1 + {\mathcal{O}} \left(
  \frac{\nu}{Q^{2 - a}}, \frac{\nu^{\frac{2}{2 - a}}}{Q^2} \right) \right],
  \,\,\, \label{ptnp} \\ 
  \ln \left[\tilde{f}_{a,{\rm NP}}
  \left(\frac{\nu}{Q},\kappa\right) \right] & \equiv & \frac{1}{1 - a} \,
  \sum\limits_{n = 1}^\infty \lambda_n(\kappa)
  \left( - \frac{\nu}{Q} \right)^n. 
\label{adep} 
\ea 
The shape function $\tilde{f}_{a,{\rm NP}}$ sums all power corrections
of the form $(\nu/Q)^n$, with unknown nonperturbative coefficients
$\lambda_n (\kappa)$, which can be formally expressed in terms of infrared 
moments of the cusp anomalous dimension $A(\as)$ as
\be
  \lambda_n(\kappa) = \frac{2}{n \, n!} \, \int_0^{\kappa^2}
  \frac{d p_\perp^2}{p_\perp^2} \left( p_\perp^2 \right)^{n/2} A \left( \as (p_\perp^2) \right)~.
\label{momA}
\ee
As explicitly indicated in \eq{ptnp}, terms of order $\nu/Q^{2 - a}$
and of order $\nu^{\frac{2}{2 - a}}/Q^2$ have been neglected. At this
level of accuracy, one finds that the only $a$-dependence of the shape
function is through an overall factor $1/(1 - a)$, which leads to the
scaling rule \cite{GSParis,CPhD,BS1}
\be
  \tilde{f}_{a, {\rm NP}} \left(\frac{\nu}{Q}, \kappa\right) = \left[
  \tilde{f}_{0, {\rm NP}} \left(\frac{\nu}{Q}, \kappa\right) 
  \right]^{\frac{1}{1 - a}}~. 
\label{rule} 
\ee 
The derivation of the scaling rule in \eq{rule} relies on two main
assumptions. First, contributions from correlations between
hemispheres are neglected, because the starting point is the NLL
resummed cross section, which describes logarithmic corrections due to
independent radiation off two back-to-back jets. In the more general
resummed formula valid to all logarithmic orders \cite{BKS1}, such
correlations are present, but they contribute only starting at NNLL
order. In addition, numerical studies indicate that inter-hemisphere
correlations do not play an important role
\cite{shapepar,Gardi:2002bg}. One can furthermore argue that correlations
between hemispheres due to particles whose decay products enter both
hemispheres become non-negligible in the same range of the parameter
$a$ where also recoil effects need to be taken into account. The
neglect of inter-hemisphere correlations is thus consistent with the
resummation. The second assumption entering the derivation of \eq{rule}
is that nonperturbative soft radiation has the same properties under
boosts as the relatively harder perturbative component. A success of
experimental tests of the scaling rule would thus show that boost
invariant dynamics dominates the differential distributions at all
scales and that coherent interjet radiation is nondominant in the
relevant range of the parameter $a$.

Of course, even if the above assumptions hold, there are further
corrections present, suppressed relative to the dominant ones, as
indicated in \eq{ptnp}. These corrections, which for thrust behave
like $\nu/Q^2$, become important only in the extreme nonperturbative
region $1 - T \sim (\LQCD/Q)^2$, and are unlikely to play a role for
phenomenology. From a theoretical point of view, it may however be of
some interest to compare the predictions of different models also for
these subleading corrections. Considering the resummation, for
example, one can keep terms neglected in \eq{rule}, and parameterize
them in terms of different integrals of the anomalous dimensions
$A(\as)$ and $B(\as)$ in the infrared region. One finds that the
pattern of subleading corrections can be characterized in terms of a
subleading shape function as follows,
\be
  \tilde{\sigma} \left(\nu, a\right) = \tilde{\sigma}_{\rm PT} 
  \left(\nu, \kappa, a\right) \, \tilde{f}_{a, {\rm NP}}
  \left(\frac{\nu}{Q},\kappa\right) \, \tilde{g}_{a, {\rm NP}}
  \left(\frac{\nu}{Q^{2 - a}},\kappa\right)~,
\label{sub}
\ee
where
\ba
  \ln \left[\tilde{g}_{a, {\rm NP}} \left(\frac{\nu}{Q^{2 - a}},\kappa
  \right) \right] & \equiv & \frac{1}{1 - a} \,
  \sum\limits_{n = 1}^\infty \overline{\lambda}_n^A (\kappa, a)
  \left( - \frac{\nu}{Q^{2 - a}} \right)^n
  \nonumber \\
  & + & \sum\limits_{n = 1}^\infty \overline{\lambda}_n^B (\kappa)
  \left( - \frac{(\nu/2)^{2/(2 - a)}}{Q^2} \right)^n~. 
\label{subsub}
\ea
The new nonperturbative parameters defining the subleading shape function 
$\tilde{g}$ are given by expressions similar to \eq{momA},
\ba
  \overline{\lambda}_n^A (\kappa, a) & = & - \frac{2}{n \, n!} \, 
  \int_0^{\kappa^2} \frac{d p_\perp^2}{p_\perp^2} \left( p_\perp^2 \right)^{(2 - a) n/2} 
  A \left( \as (p_\perp^2) \right)~,
\nonumber \\
  \overline{\lambda}_n^B (\kappa) & = & \frac{1}{n!} \, 
  \int_0^{\kappa^2} \frac{d p_\perp^2}{p_\perp^2} \left( p_\perp^2 \right)^{n/2} 
  B \left( \as (p_\perp^2) \right)~.
\label{subpar}
\ea 
We notice that for both kinds of subleading contributions there is no
simple scaling behavior with $a$. Furthermore, both contributions
are suppressed at large $\nu$ (that is, small $\tau_a$), with an
increasing degree of suppression as $a$ grows large and negative:
specifically, subleading power corrections appear as functions only of
the combination $\nu/Q^{2 - a}$, a feature that will also be
found in the  DGE formalism.

In the following, we will study leading and subleading power
corrections by means of DGE, and we will compare the results obtained
with those arising directly from the resummation described above.

\section{Dressed gluon exponentiation for angularities} 
\label{DGEsec}

Dressed gluon exponentiation begins with a conventional renormalon
analysis of the given event shape. One computes the characteristic
function of the dispersive approach~\cite{dispers}, that is, the
contribution to the cross section of a single gluon dressed with an
arbitrary number of quark bubbles. This is referred to as the ``single
dressed gluon'' (SDG) cross section. One proceeds by identifying the
terms in the characteristic function which contribute to logarithmic
behavior in the two-jet limit. These terms can be exponentiated,
resumming all contributions where any number of dressed gluons is
emitted without interfering. DGE reproduces 
resummation at NLL, provided the running coupling is defined in the
bremsstrahlung scheme. Furthermore, all subleading logs are accounted
for in the large $\beta_0$ (large $n_f$) limit. This way, DGE can detect 
 the factorial growth of the coefficients of subleading logarithms,
and provide methods to gauge the range of applicability of
conventional Sudakov resummations. Furthermore, given an explicit
representation of the singular behavior of perturbation theory, one
may pick a definite prescription to deal with the resulting
ambiguities.

\subsection{Angularities with a single dressed gluon} 
\label{SDGsec}

Summing up infinitely many bubbles in the gluon propagator in the
inclusive approximation is equivalent to performing a calculation
with an off-shell gluon with virtuality $k^2$, and replacing the
coupling with an effective, ``timelike coupling''
\cite{Ball:1995ni}. One then expresses the SDG cross section by
\be
  \left. \frac{1}{\sigma_{\rm tot}} \frac{d \sigma}{d \tau_a} 
  \right|_{\SDG} = \frac{C_F}{2 \beta_0} \int_0^1 \frac{d
  \epsilon}{\epsilon} \bar{A}_{\eff}\left(\epsilon Q^2 \right)
  \dot{\mathcal{F}}(\tau_a,\epsilon)~.
\label{SDG}
\ee
Here $\dot{\mathcal{F}}$ is the derivative with respect to the gluon
virtuality $\epsilon = k^2/Q^2$ of the ``characteristic function''
$\mathcal{F}$,
\ba
  \dot{\mathcal{F}}(\tau_a, \epsilon) & = & - \epsilon
  \frac{\partial}{\partial \epsilon} {\mathcal{F}}(\tau_a,\epsilon)~,
  \nonumber \\
  {\mathcal{F}}(\tau_a, \epsilon) & = & \int d x_1 d x_2 
  \left| M(x_1, x_2, \epsilon) \right|^2 \delta\left( \tau_a -
  \tau_a(x_1, x_2, \epsilon) \right)~. 
\label{charf}
\ea
The squared matrix element $\left| M \right|^2$ for the emission of a
gluon of virtuality $k^2$ in the process $\gamma^* \rightarrow q
\bar{q} g$, without coupling and color prefactors, is given by
\be
  \left| M(x_1, x_2, \epsilon) \right|^2 = \frac{(x_1 + \epsilon)^2 +
  (x_2 + \epsilon)^2}{(1 - x_1)(1 - x_2)} - \frac{\epsilon}{(1 - x_1)^2} -
  \frac{\epsilon}{(1 - x_2)^2}~. 
\label{matelf}
\ee
$x_1$ and $x_2$, as usual, are the energy fractions of the quark and
antiquark in the center-of-mass frame, and one can define $x_3 = 2 -
x_1 - x_2$, the gluon energy fraction. When the gluon is ``massive'',
the limits of phase space are given by
\ba
  x_1 + x_2 & \geq  & 1 - \epsilon 
  \nonumber \\
  (1 - x_1)(1 - x_2) & \geq & \epsilon~. 
\label{PSeps}
\ea
The final ingredient in \eq{SDG} is the ``timelike coupling'' 
$\bar{A}_{\eff}$, which is typically expressed in terms of a Borel 
representation as
\be
  \bar{A}_{\eff} \left( Q^2 \right) = \int_0^\infty d u \left(
  \frac{Q^2}{\LQCD^2} \right)^{- u} \frac{\sin \pi u}{\pi u}
  \bar{A}_B (u)~. 
\label{Aeff}
\ee

We now need to specify a suitable generalization of the massless
definition of the event shape, \eq{barfdef}, for the case in which the
emitted gluon has nonvanishing virtuality. Several observations are
helpful in deciding how to perform this generalization.  First, all
logarithmic contributions to the SDG cross sections stem from the
region where the gluon is either soft or collinear. Contributions from
the region where the gluon is relatively hard do not give logarithmic
enhancements, and the exact location of the boundary between the
region of phase space where one of the quarks is dominant and the
region where the gluon dominates is unimportant. We can thus
concentrate on the phase space region where the thrust axis is the
quark momentum, and $x_1$ the largest energy fraction. Contributions
in which the antiquark momentum dominates can be obtained by symmetry.
Recall the expression for $\tau_a$ in the case of massless partons,
in the region where the quark has the largest energy,
\be
  \tau_a(x_1, x_2) = \frac{(1 - x_1)^{1 - a/2}}{x_1} \left[
  (1 - x_2)^{1 - a/2} (x_1 + x_2 - 1)^{a/2} + (x_1 \leftrightarrow 
  x_2) \right]~. 
\label{x1weps0}
\ee
We would like our definition to reduce to \eq{x1weps0} as $\epsilon
\to 0$.  Furthermore, we require that for $a = 0$ the definition
should reduce to the massive definition of the thrust, as used for
example in \cite{Gardi:2000yh,Gardi:2001ny,Gardi:2002bg,Gardi:1999dq}.
In the phase space region at hand, this is simply $\tau_0 (x_1, x_2,
\epsilon) = 1 - x_1$. Finally, as we will see, working analytically
for generic $a$ and with a massive gluon generates rather intricate
expressions, so we must keep the definition as simple as possible in
order to be able to perform the necessary integrations.

Keeping these criteria in mind, we define the angularity with an
off-shell gluon, in the region where the thrust axis is given by the
quark momentum, as
\ba
  \tau_a(x_1, x_2, \epsilon) & = & \frac{(1 - x_1)^{1 - a/2}}{x_1} 
  \left[(1 - x_2 - \epsilon)^{1 - a/2} (x_1 + x_2 - 1 + \epsilon)^{a/2}
  \right. \nonumber \\ & & \left. \qquad \qquad \quad + 
  (x_1 + x_2 - 1 + \epsilon)^{1 - a/2} (1 - x_2 - \epsilon)^{a/2} \right]~. 
\label{x1weps}
\ea
Of course, other choices satisfying our criteria are possible. It can
be shown, however, that different treatments of the gluon mass alter
the value of $\tau_a$ by terms that are suppressed by higher powers of
the weight, roughly by a factor of $\tau_a^{1-a}$. Thus, they do not
change logarithmically enhanced contributions. Once a definite
prescription to include massive partons is chosen, the predictions
within the DGE formalism are unambiguous. It should be kept in mind,
in any case, that a comparison to experiment requires a detailed
analysis along the lines of Ref.~\cite{Salam:2001bd} for the inclusion
of hadron mass effects.

To proceed, it is useful to change integration variables from
$x_1, x_2$ to
\ba
  \zeta & = & 1 - x_1~,
  \nonumber \\
  \xi & = & \frac{x_1 + x_2 - 1 + \epsilon}{1 - x_2 - \epsilon}~. 
\label{changevar}
\ea
In terms of these variables, the characteristic function can be
written as
\ba
  {\mathcal{F}}(\tau_a,\epsilon) & = &
  \int\limits_{\epsilon}^{\sqrt{\epsilon}} d \zeta
  \int\limits^{\frac{\zeta}{\epsilon} - 1}_{\frac{\zeta}{1 - 2 \zeta}} 
  d \xi \left[ \frac{(1 - \zeta + \epsilon)^2 (1 - \zeta)}{\zeta (1 - 
  \zeta + \epsilon(1 + \xi))(1 + \xi)}
  \right. \nonumber\\
  & & \quad \left. + \quad \frac{(\xi + \zeta)^2 (1 - \zeta)}{\zeta 
  (1 - \zeta + \epsilon (1 + \xi))(1 + \xi)^3} - \frac{\epsilon (1 - 
  \zeta)}{(1 - \zeta + \epsilon(1 + \xi))^2} \right. \nonumber \\
  & & \quad \left. - \quad \frac{\epsilon (1 - \zeta)}{\zeta^2 (1 + \xi)^2}
  \right] \delta \left(\tau_a - \frac{\zeta^{1 - a/2}}{1 + \xi} \xi^{a/2}
  (1 + \xi^{1 - a}) \right)~. 
\label{charF}
\ea
Two comments are in order. First, the limits of integration do
not correspond to the full phase space, but only to the region which
generates logarithmically enhanced contribution, that is, the region
in which the gluon is either soft or collinear to the quark. With a
massive gluon, the collinear limit corresponds to $x_1 = 1 -
\epsilon$, $x_2 = 0$, whereas the soft limit is given by $x_1 = x_2 =
1 - \sqrt{\epsilon}$. These two values set the limits of the $\zeta$
integration. The soft boundary of phase space for intermediate values
of $\zeta$ is given by $x_2 \leq 1 - \epsilon/(1 - x_1)$,
corresponding to the upper limit of the $\xi$ integration. The Sudakov
region thus corresponds to values of $\xi$ close to the upper boundary
of integration. The second observation is that, with this choice for
angularity in the presence of a massive gluon, and in these variables,
the $\delta$-function defining the curves of constant angularity in
\eq{charF} does not depend on $\epsilon$, a feature that will be
exploited below.

\subsection{Exponentiation and Borel representation}

The SDG cross section summarizes the probability that one dressed
gluon is emitted. Upon summing over probabilities that many such
dressed gluons are emitted independently, one obtains the
exponentiated expression \cite{Gardi:2001ny,Gardi:2002bg}
\be
  \left. \frac{1}{\sigma_{\rm tot}} \tilde{\sigma}(\nu, a)
  \right|_{\DGE} = \exp \left[ \int_0^\infty d \tau_a \left( e^{- \nu
  \tau_a} - 1 \right) \left( \left. \frac{1}{\sigma_{\rm tot}} 
  \frac{d \sigma}{d \tau_a} \right|_{\SDG} \right)
  \right]~. 
\label{DGE}
\ee
Here we have extended the integration region beyond the support of the
SDG cross section. This, however, does not change the result in the
region of interest, at small $\tau_a$ or equivalently at large
$\nu$. Inserting \eq{SDG} and \eq{Aeff} into \eq{DGE}, one may write
down an explicit Borel representation for the exponent, although at
this point one is still dealing with a five-fold integral. One finds
\be
  \ln \left[\frac{1}{\sigma_{\rm tot}} \tilde{\sigma}(\nu, a) \right]_{\DGE} \!\!\!
  = \frac{C_F}{2 \beta_0} \int\limits_0^\infty d u 
  \left( \frac{Q^2}{\LQCD^2} \right)^{-u} \frac{\sin \pi u}{\pi u} 
  \bar{A}_B(u) \int\limits_0^\infty \frac{d \epsilon}{\epsilon} 
  \epsilon^{- u} \int\limits_{\tau_a^{\rm co}}^{\tau_a^{\rm s}} 
  d \tau_a \left(e^{- \nu \tau_a} - 1 \right) 
  \dot{\mathcal{F}}(\tau_a, \epsilon).
\label{genexp}
\ee
To exchange the order of integration, as done in \eq{genexp}, it has
been necessary to evaluate the limiting values of
angularity in the phase space region of interest, for a fixed gluon
virtuality $\epsilon$.  One finds that for any value of $a$ the
collinear limit of angularity is given by $\tau_a^{\rm co} = \epsilon
+ {\cal O} (\epsilon^2)$, while the soft limit is $\tau_a^{\rm s} =
\sqrt{\epsilon} + {\cal O} (\epsilon)$ (notice that the corrections
neglected here vanish for $a = 0$). As shown below, however, the exact
form of these limits is not important: only the values of the energy
fractions of the quark and of the gluon at the soft and collinear
boundaries of phase space are relevant, and they are independent of
the weight $\tau_a$, and solely dependent on the kinematics.

The advantage of writing the exponent as in \eq{genexp} is that it is
now possible, before performing phase space integrals, to take the
derivative with respect to $\epsilon$ of \eq{charF}, thanks to the
fact that the $\delta$-function defining $\tau_a$ does not depend on
$\epsilon$ with our choice of variables. One may then use the 
$\delta$-function to perform trivially the $\tau_a$ integral.  Discarding terms
that do not contribute Sudakov logarithms, we can write the Borel
representation of the exponent as
\be
  \ln \left[\frac{1}{\sigma_{\rm tot}} \tilde{\sigma} (\nu, a)
  \right]_{\DGE} = \frac{C_F}{2 \beta_0} \int_0^\infty d u \left(
  \frac{Q^2}{\LQCD^2} \right)^{- u} \frac{\sin \pi u}{\pi u}
  \bar{A}_B(u) \, B (u, \nu, a)~, 
\label{Borelln}
\ee
with the Borel function $B(u,\nu,a)$ given by
\ba
  B (u, \nu, a) \hspace{-5pt} & \equiv & \hspace{-5pt} 
  \int\limits_0^\infty d \epsilon \,\epsilon^{-u} 
  \int\limits_{\epsilon}^{\sqrt{\epsilon}} d \zeta \left\{ 
  \int\limits_{0}^{\zeta/\epsilon - 1} d \xi
  \frac{1}{\zeta^2 (1 + \xi)^2} \left[ \exp \left( - \nu \zeta^{(2 - a)/2}
  \frac{\xi^{a/2}}{1 + \xi} (1 + \xi^{1 - a}) \right) - 1 \right] \right.
  \nonumber \\ & + & \hspace{-8pt} \left. \left(\frac{2}{\epsilon \zeta} - 
  \frac{2}{\zeta^2} \right) \left[ \exp\left\{ - \nu \, \epsilon \, 
  \zeta^{-a/2} \left( \frac{\zeta}{\epsilon} - 1 \right)^{a/2}
  \left(1 + \left( \frac{\zeta}{\epsilon} - 1 \right)^{1 - a} \right) 
  \right\} - 1 \right] \right\}.
\label{Buh}
\ea
The factors of $- 1$ in \eq{Buh} are due to virtual corrections. They
contribute terms independent of $a$, and are thus identical to the thrust
\cite{Gardi:2001ny,Gardi:2002bg}. They do not need to be considered
anew. We will now concentrate on the contributions associated with
real gluon emission, labelled in the following by an additional 
subscript, $B_R (u, \nu, a)$.

It is possible to arrive at a one-dimensional integral representation
of $B_R (u, \nu, a)$. First we perform the integration over $\zeta$ in
the first term of \eq{Buh}, and change variables from $\zeta$ to
$\xi = \zeta/\epsilon - 1$ in the second term.  Finally, integrating over
$\epsilon$ results in the following expression
\ba
  B_R(u, \nu, a) & = & - \frac{1}{u} \, \int\limits_0^\infty d \xi 
  \, (1 + \xi)^{2 u} \, \left[ \frac{2}{1 + \xi} - \frac{2}{(1 + 
  \xi)^2} + \frac{1}{1 - u} \frac{1}{(1 + \xi)^3} \right]
  \nonumber \\ & & \times \mbox{}_1F_1 \left[ - \frac{2 u}{2 - a}; 
  1 - \frac{2 u}{2 - a}; - \nu \xi^{a/2} \frac{1 + \xi^{1 - 
  a}}{\left(1 + \xi \right)^{2 - a/2}} \right] + {\cal O} 
  \left(\frac{1}{\nu} \right) \, ,
\label{final1}
\ea
where $\mbox{}_1F_1$ is the confluent hypergeometric function, also
known as Kummer's function of the first kind, defined by \cite{GR}
\be
  \mbox{}_1F_1 \left( \alpha; \beta; z \right) \equiv 
  \sum\limits_{k = 0}^\infty \, \frac{(\alpha)_k}{(\beta)_k} \, 
  \frac{z^k}{k!}~,
\ee
where the Pochhammer symbol is defined as
\be
  (\alpha)_k \equiv \alpha (\alpha + 1) \dots (\alpha + k - 1) =
  \frac{\Gamma(\alpha + k)}{\Gamma(\alpha)}~. 
\label{poch}
\ee
Clearly, \eq{final1} cannot be evaluated directly. Nevertheless, in
the physically interesting limits, the soft and collinear region of
phase space, results in closed form can be obtained, and from these
results the form of power corrections can be inferred. Before
discussing the results for general $a$, however, we pause to note that
we can test \eq{final1} by considering the case of the thrust, $a = 0$, where the
answer is known~\cite{Gardi:2001ny,Gardi:2002bg}.  For $a = 0$,
indeed, \eq{final1} can be explicitly integrated. From the resulting
expressions one can see that only the factor $2/(1 + \xi)$ in the
square bracket of \eq{final1} contributes in the soft limit, whereas
the remaining contributions are of purely collinear origin. The final
result for $B_R (u, \nu, 0)$ is
\be
  B_R(u, \nu, 0) = \nu^{2 u} \, \Gamma(- 2 u) \frac{2}{u} - \nu^u \,
  \Gamma(- u) \left(\frac{2}{u} + \frac{1}{1 - u} + \frac{1}{2 - u} 
  \right)~,
\label{thrustu}
\ee
up to corrections of order $1/\nu$. Our result is in complete
agreement with Refs.~\cite{Gardi:2001ny,Gardi:2002bg}. Soft
contributions produce singularities at half-integer values of $u$,
corresponding to power corrections of the form $(\LQCD \nu/Q)^p$;
collinear contributions have poles at integer $u$, giving power
corrections behaving as $(\LQCD^2 \nu/Q^2)^p$. We note also that the
result in \eq{thrustu} is infrared safe at leading power. In other
words, the poles at $u = 0$ (corresponding to $Q^0$) cancel between
the soft and the collinear contributions. We will verify
this key property of infrared safety also for general $a$.

\section{The pattern of exponentiated power corrections}
\label{papo}

\subsection{Power corrections from soft radiation}
\label{paposo}

We now turn to the soft contribution for general $a$, which is the
source of the scaling in \eq{rule}. In the region of phase space
corresponding to soft emission one has $\zeta \sim \sqrt{\epsilon}$,
so that $\xi$ is large, behaving as $1/\sqrt{\epsilon}$. In this
region, we will first evaluate \eq{final1} by making use of an
asymptotic expansion for the Kummer function. Then we will go back to
the characteristic function ${\mathcal{F}}(\tau_a, \epsilon)$ and
evaluate it in the same limit, to trace the contributions that display
scaling back to the matrix element.

To treat the soft limit, we can use the following integral
representation of the confluent Kummer function \cite{wolfram},
\be
  \mbox{}_1 F_1 (\alpha; \beta; z) = \frac{\Gamma(\beta)}{\Gamma(\alpha)
  \Gamma(\beta - \alpha)} \int_0^1 d t \, e^{z t} \, t^{\alpha - 1}
  (1 - t)^{- \alpha + \beta - 1}~.
\label{wolfie}
\ee
We then see that the complicated argument of the Kummer function in
\eq{final1} appears only in the exponent, and can be expanded for
large $\xi$ and $\nu$, with $\nu/\xi$ held fixed. One finds
\be
  \exp \left[ - \nu \, t \, \xi^{a/2} \frac{1 + \xi^{1 - 
  a}}{\left(1 + \xi \right)^{2 - a/2}} \right] = \exp
  \left[ - \frac{\nu \, t}{\xi} \right] + \dots~.
\label{expexp}  
\ee 
The soft contribution then simplifies considerably, and becomes
\ba
  B_R^{\rm soft} \left(u, \nu, a \right) & = & \frac{2}{2 - a}
  \, \int\limits_0^\infty d \xi \, \int\limits_0^1 d t \, e^{- \nu \, t/\xi}
  \, t^{- \frac{2 u}{2 - a} - 1} \, (1 + \xi)^{2 u} 
  \nonumber \\ & & \times \, \,
  \left[ \frac{2}{1 + \xi} - \frac{2}{(1 + \xi)^2} + \frac{1}{1 - u}
  \frac{1}{(1 + \xi)^3} \right] + \,
  {\mathcal{O}} \left(\frac{1}{\nu} \right) \, . 
\label{presoft}
\ea
Since we are interested in the large $\xi$ limit, clearly at leading
power only the first term in the bracket of \eq{presoft} contributes,
while the remaining two terms are suppressed by powers of $1/\nu$. The
integral can be performed, and one finds that the final answer for the
soft contribution is
\ba
  B_R^{\rm soft} \left(u, \nu, a \right) = \frac{1}{1 - a} \,
  \nu^{2 u} \, \Gamma(- 2 u) \, \frac{2}{u} +
  {\mathcal{O}}\left( \frac{1}{\nu} \right) \, . 
\label{finalsoft}
\ea
Comparing this with the case for $a = 0$, \eq{thrustu}, we see that we
recover exactly the prediction of \eq{rule}. We have thus confirmed
the result of \cite{BS1}, obtained there by analyzing the resummed
expression.

\subsection{An alternative derivation of the scaling rule}
\label{alpaposo}

To perform the integrations leading to \eq{finalsoft} for generic $a$,
it has been necessary to make use of some special function
technology. This technology may obscure the physics underlying the
simple scaling, and indeed the result of \eq{finalsoft} may appear
somewhat surprising, in view of the fact that, for example,
\eq{presoft} still has an overall factor of $2/(2 - a)$, corresponding
to the scaling of leading logarithms but not to the scaling of power corrections.
It is thus instructive to recover the result by going back to the
characteristic function, \eq{charF}, and identifying the origin of the
terms responsible for the leading soft contribution.

The first observation to simplify \eq{charF} concerns phase space. As
observed before, the Sudakov region corresponds to a neighborhood of
the upper limit of the $\xi$ integration range, whereas the lower
limit does not contribute to logarithmic enhancements and can be
changed.  To simplify the calculation, we can for example take as a
lower limit $\xi = 0$, as was already done in \eq{Buh}. The expression
for ${\mathcal{F}}$ becomes then slightly more manageable by changing
variables from $\xi$ to $\omega = 1 + \xi$. One finds
\ba
  {\mathcal{F}}(\tau_a, \epsilon) & = &
  \int\limits_{1}^{1/\sqrt{\epsilon}} d \omega
  \int\limits_{\epsilon \, \omega}^{\sqrt{\epsilon}} d \zeta \, \,
  \frac{1 - \zeta}{\zeta} \left[ \frac{(1 - \zeta + \epsilon)^2}{\omega 
  (1 - \zeta + \epsilon \omega)}  + \frac{(\zeta + \omega - 1)^2}{\omega^3
  (1 - \zeta + \epsilon \omega)} 
  \right. \nonumber \\ && \, \, - \, \left. 
  \frac{\epsilon}{\zeta \omega^2} - \frac{\epsilon \zeta}{(1 - 
  \zeta + \epsilon \omega)^2} \right] \delta \left( \tau_a - \zeta^{1 - 
  a/2} f_a(\omega) \right)~,
\label{alcharF}
\ea
where we define
\be
  f_a(\omega) = \left(\omega - 1\right)^{a/2} \, \frac{1 + 
  (\omega - 1)^{1 - a}}{\omega}~.
\label{fa}
\ee
Notice that $f_0 (\omega) = 1$, which of course simplifies things
considerably for the thrust. For any $a$, the $\zeta$ dependence in
the $\delta$-function is sufficiently simple to be used to perform the
$\zeta$ integration, yielding a rather cumbersome expression, which
however has some interesting features. One finds
\ba
  {\mathcal{F}}(\tau_a, \epsilon) & = & \frac{2}{2 - a} \, 
  \frac{1}{\tau_a} \int\limits_{1}^{1/\sqrt{\epsilon}} 
  \frac{d \omega}{\omega} \, \left( 1 - \overline{\zeta}_a \right) \,
  \left[ \frac{(1 - \overline{\zeta}_a + \epsilon)^2}{
  (1 - \overline{\zeta}_a + \epsilon \omega)}  + \frac{(\overline{\zeta}_a + 
  \omega - 1)^2}{\omega^2 (1 - \overline{\zeta}_a + \epsilon \omega)} 
  \right. \nonumber \\ && \, \, - \, \left. 
  \frac{\epsilon}{\overline{\zeta}_a \omega} - \frac{\epsilon 
  \overline{\zeta}_a \,\omega}{(1 - \overline{\zeta}_a + \epsilon \omega)^2} 
  \right] \, \theta \left(\overline{\zeta}_a - \epsilon \, \omega \right) 
  \, \theta \left( \sqrt{\epsilon} - \overline{\zeta}_a \right)~,
\label{cumber}
\ea
where
\be
  \overline{\zeta}_a = \left( \frac{\tau_a}{f_a (\omega)} 
  \right)^{\frac{2}{2 - a}} = \frac{\left( \omega \, \tau_a 
  \right)^{\frac{2}{2 - a}} \left( \omega - 1 \right)^{- \frac{a}{2 - a}} 
  }{\left[1 + (\omega - 1)^{1 - a} \right]^{\frac{2}{2 - a}}}~.
\label{zea}
\ee
Several observations help disentangling \eq{cumber}. First, the
second $\theta$-function is not relevant to the logarithmic behavior
and can be neglected. Its effect would be to split the integration
range for $\omega$ in two subintervals, however all leading
contributions come from neighborhoods of the upper and lower limits of
integration. A second important point is the fact that the leading
singularity in $\tau_a$ is now explicitly factored out. One can then
evaluate the integral to leading power in $\tau_a$. The constraint
imposed by the first $\theta$-function can also be considerably
simplified. The integral, in fact, has support on the region defined
by
\be
  \frac{\tau_a}{\epsilon^{1 - a/2}} \geq \left( \frac{\omega}{\omega - 1}
  \right)^{-a/2} \left[ 1 + (\omega - 1)^{1 - a} \right]~.
\label{suppo}
\ee
This constraint cannot be solved exactly for $\omega$. One notices,
however, that in the physical region, $\epsilon \leq \tau_a \leq
\sqrt{\epsilon}$, and for $a \leq 0$, the left hand side of \eq{suppo}
is a parametrically large number. Within the integration range, on the
other hand, the right hand side becomes large only near the boundaries,
as $\omega \to \infty$ or as $\omega \to 1$. One can then solve the
$\theta$-function constraint in these two limits, obtaining
respectively
\be
  \omega < \omega_+ \equiv \left( \frac{\tau_a}{\epsilon^{1 - a/2}} 
  \right)^{\frac{1}{1 - a}} \, , \qquad \omega \to \infty~;
\label{lao}
\ee
and
\be
  \omega > \omega_- \equiv 1 + \frac{\epsilon^{1 - 2/a}}{\tau_a^{- 
  2/a}} \, \, , \qquad \omega \to 1~.
\label{smo}
\ee
The result of these manipulations is that the integration region in
\eq{cumber} shrinks at both boundaries, with both integration limits
now dependent on $\tau_a$. Since soft contributions arise from the
region of large $\omega$, we will concentrate on \eq{lao},
although similar arguments could be used with \eq{smo}. This will be
sufficient to recover the scaling rule.

To complete the calculation, we must now approximate the matrix
element. To this end, note that for large $\omega$ one can approximate
\eq{zea} by
\be
  \overline{\zeta}_a = \tau_a^{\frac{2}{2 - a}} \omega^{\frac{a}{2 - 
  a}} \ll 1~.
\label{appzea}
\ee
One sees that $\overline{\zeta}_a$  is a
small number in the relevant range of $a$, 
and we can expand the integrand of \eq{cumber} around
$\overline{\zeta}_a = 0$. The resulting soft approximation of the
characteristic function is
\be
  \left. {\mathcal{F}}(\tau_a, \epsilon) \right|_{\rm soft} =
  \frac{2}{2 - a} \, \frac{1}{\tau_a} \, \int\limits_{}^{\omega_+} 
  \frac{d \omega}{\omega} \left[ 2 - \frac{2}{\omega} + 
  \frac{1}{\omega^2} - \frac{\epsilon}{\left( \tau_a \omega
  \right)^{\frac{2}{2 - a}}}\right]~.
\label{sofF}
\ee
The integration can now be performed and compared with the result for
the thrust. At large $\omega$ the dominant contribution clearly comes
from the first term in the square bracket. Comparing with the case $a
= 0$, we see in fact that, while all terms contribute to the Sudakov
limit, the last three terms give subleading corrections associated
with collinear radiation. This remains true for generic $a$, as noted
in the following subsection. Concentrating on the first term, we can
finally see how the scaling rule arises in the context of DGE. 
The only term in the cross section contributing in the soft
limit is, as might be expected, the logarithmic integral over the
bremsstrahlung gluon spectrum. The gluon mass then acts as an infrared
cutoff on this integral. The power of $\epsilon$ in the upper limit of
integration $\omega_+$ is precisely the one required to cancel the
overall factor of $2/(2 - a)$, and to replace it with the scaling factor
of \eq{rule}. This precise power arises uniquely from the definition
of angularity, as expressed by the $\delta$-function in \eq{alcharF},
and can easily be traced back to the exponential weight given to
pseudorapidity in \eq{barfdef}. The final result in the soft limit is
simply
\be
  \left. {\mathcal{F}}(\tau_a, \epsilon) \right|_{\rm soft} =
  - \, \frac{1}{1 - a} \, \frac{2}{\tau_a} \, \ln(\epsilon)~,
\label{short}
\ee
where subleading collinear contributions and terms independent of
$\epsilon$, which do not contribute to the logarithmic behavior, have
been omitted. Power corrections of collinear origin can be treated
similarly, giving results consistent with the ones outlined in the
following subsection.

\subsection{Power corrections from collinear radiation}
\label{papoco}

At this point we have various methods at our disposal to analyze power 
corrections of collinear origin. We can for example complete the analysis
of \secn{alpaposo}, including the effects of the lower limit of integration,
and then inserting the results in the Borel exponent. Alternatively, we can 
go back to \eq{final1} and study it in this limit.  Collinear power corrections
arise at large $\nu$, but small $\xi$. In this limit, \eq{final1} yields
\ba
  B_R^{\rm coll.} (u, \nu, a) & = & \frac{2}{2 - a}
  \Gamma \left(- \frac{2 u}{2 - a} \right) \nu^{\frac{2 u}{2 - a}}
  \, \int\limits_0^\infty d \xi \, \xi^{\frac{a}{2 - a} u}
  (1 + \xi)^{- \frac{a}{2 - a} u} (1 + \xi^{1 - a})^{\frac{2 u}{2 - a}} 
  \,\nonumber \\ & & \,\,\, \times \left[ \frac{2}{1 + \xi} - 
  \frac{2}{(1 + \xi)^2} + \frac{1}{1 - u} \frac{1}{(1 + \xi)^3} \right] +
  {\mathcal{O}}\left( \frac{1}{\nu} \right) \, .
\label{finalco}
\ea
We see immediately that collinear contributions are suppressed by
non-integer powers compared to the leading soft piece, as found above
in the derivation of \eq{rule}. This non-integer power behavior can
also be seen from \eq{finalco}, which has singularities at non-integer
$u$ due to the $\Gamma$-function, and further singularities due to the
integration over $\xi$, which can be readily performed for any
specific value of $a$. For generic $a$, the leading poles on the
positive real $u$ axis can also be determined by studying the
integrand of \eq{finalco} in the limits of large and small $\xi$.
Determining the full analytic structure of \eq{finalco} as a function
of $a$ is however considerably more difficult. In the case when
$a$ is a rational number, we have been able to obtain a closed
expression in terms of generalized hypergeometric functions $\mbox{}_p
F_q$.  The detailed expressions are listed in Appendix A. Since all
methods give consistent results, we will just summarize here the
structure of poles on the positive real axis of the Borel variable
$u$, and we will outline the corresponding pattern of power
corrections.

First, let us note that our complete result for the Borel
function $B$ is infrared safe, as it must. In fact, as was the case
for the thrust, the poles at $u = 0$ cancel between soft and collinear
contributions.  To see this, note for example that the intricate
expression in \eq{mintcoend2} simplifies at $u = 0$ to 
\be
  B_R^{\rm coll.} (u \sim 0, \nu, a) =
  \nu^{\frac{2}{2 - a} u} \, \Gamma \left(- \frac{2}{2 - a} u \right)
  \frac{4}{2 - a} \Gamma \left(- \frac{2 (1 - a)}{2 - a} u \right)~,
\label{calpo}
\ee
where we have suppressed terms nonsingular at $u=0$. 
This clearly cancels against the contribution of \eq{finalsoft} at $u
= 0$, as expected.

Next, we note that \eq{finalco} appears to have an explicit pole at 
$u = 1$, which would correspond to a correction of order ${\mathcal{O}}
\left(\nu (\LQCD/Q)^2\right)$. This pole, however, is cancelled by the 
explicit factor of $\sin \pi u$ in \eq{genexp}, as was the case for the 
thrust (where power corrections of this form are present only as an 
effect of the $\Gamma$-functions in \eq{thrustu}).

The general structure of poles in $u$, for rational $a$, can be
deduced from \eq{mintcoend2} in Appendix A. There are several infinite
towers of poles.  From \eq{finalco} we can directly read off the first
pole in each tower. We find singularities at
\ba
  u & = & u_1 \, = \, \frac{2 - a}{2}~, \nonumber \\
  u & = & u_2 \, = \, - \, \frac{2-a}{a}~, \label{copole}  \\
  u & = & u_3 \, = \, \frac{1}{2} \, \frac{2 - a}{1 - a}~. 
\nonumber
\ea
Recall that $a \leq 0$ in the range of validity of our approach. The
crucial fact here is that all these singularities are accompanied by a
common factor of $\nu^{2 u/(2 - a)}$, as seen in \eq{finalco}. When
combined with the locations of the poles in \eq{copole}, we see that
all collinear power corrections are expressed in terms of a single
combination of $\nu$ and of the scale $Q$, namely $\nu (\LQCD/Q)^{2 -
a}$, possibly further raised to a non-integer power. This result agrees
with the estimate extracted for the resummation, Eqs. (\ref{sub})-(\ref{subsub}), 
although the detailed
pattern of subleading singularities is different. We conclude that
collinear power corrections are suppressed as predicted from resummation, and are
expected to become important only for extreme values of the angularity
$\tau_a$, $\tau_a \sim {\cal O} (\LQCD/Q)^{2 - a}$. The suppression grows
as $a$ becomes large and negative, although numerically the effect could
be partly compensated by the fact some of the corrections may be
further raised to small non-integer powers, of order $1/(1 - a)$.

It is also important to note that the intricate structure of towers of
subleading poles arising from \eq{mintcoend2} is not as
model-independent as the leading poles connected to soft radiation.
For example, a different choice for the massive definition of
angularity, \eq{x1weps}, might kinematically generate  non-integer power
corrections of comparable size. We emphasize, in any case, that the
leading structure expressed by \eq{finalsoft} is
unaffected. Furthermore, we believe that the parametric dependence of
collinear power corrections on the ratio $\nu (\LQCD/Q)^{2 - a}$,
which is found via resummation, and confirmed by our DGE analysis, is
a stable feature, suggesting that the scaling rule should hold with
increasing precision for negative $a$.

\section{Conclusions}
\label{conclu}

In this work we have verified that the universal scaling of the
leading power corrections within the class of angularities, discovered
in Ref.~\cite{BS1} from soft gluon resummation, is unchanged when, in
addition to large Sudakov logarithms, also renormalon enhancements are
taken into account. Furthermore, we have been able to determine the
form of subleading corrections within a specific scheme to treat
massive partons. These corrections, of collinear origin, are
suppressed by non-integer powers of the hard scale relative to the
leading ones that originate from soft radiation. Different models of
these collinear corrections further suggest that the relative
suppression grows as the parameter $a$ becomes large and negative.  
A detailed phenomenological study along the lines of
Refs. \cite{Gardi:2001ny,Gardi:2002bg} could help to further constrain
the nonleading effects, and would be necessary for a comparison with
experiment. We defer such a study to future work, due to the lack of
corresponding experimental data.

Assuming subleading corrections are negligible, as indeed our results imply,
the scaling allows to predict the distributions of the whole class of
angularities, including nonperturbative corrections, in terms of a
single shape function, which can in principle be determined from data
by considering a specific choice of the parameter $a$, for example the
thrust, $a = 0$. Since there are no free parameters, an experimental
determination of distributions for other values of $a$ would certainly
give valuable information about the properties of nonperturbative
corrections. Also, since the perturbative cross section has a scaling
behavior with $a$ different from the non-perturbative shape function,
comparison of theory and experiment might help to disentangle
corrections due to missing higher-order perturbative information from
power correction effects. An experimental test of
scaling would determine to what extent the boost invariance of soft
radiation in the two-jet limit, which is established at the
perturbative level, also applies to nonperturbative effects. We hope
therefore that an analysis of experimental data for the class of
angularities will be made in the near future.

\begin{appendix}

\section{Evaluation of the collinear contribution}
\label{appA}

Here we sketch the evaluation of the collinear contribution to the DGE
cross section for rational $a$, that is, we consider the case
\be
1- a = \frac{p}{q},\,\quad p,q \mbox{ positive integer}, p \geq q.
\ee 

This is sufficient for our purposes, since any irrational number can
always be well approximated by a rational one.

The collinear limit is equivalent to the limit in which $\nu
\rightarrow \infty$ faster than $\xi$ (although $\xi$ varies between 0
and $\infty$). In this limit the collinear contribution is given by Eq. (\ref{finalco}).
In the following we will abbreviate Eq. (\ref{finalco}) by
\be
B_R^{\rm coll.}(u,\nu,a) = \frac{2}{2-a} \Gamma\left(- \frac{2 u}{2-a}
\right) \nu^{\frac{2 u}{2-a}} \sum\limits_{m=1}^3 c_m {\mathcal{I}}^m,
\label{abbrev}
\ee
where
\ba
{\mathcal{I}}^m & = & \int\limits_0^\infty d \xi \,\xi^{\frac{a}{2-a}
u} (1+\xi)^{-\frac{a}{2-a}u-m} (1+\xi^{1-a})^{\frac{2 u}{2-a}}
\label{Idef} \\ c_1 & = & 2,\qquad \quad c_2 = -2, \qquad \quad c_3 =
\frac{1}{1-u}.
\ea

We use the Mellin-Barnes representation of
$(1+\xi^{1-a})^{2u/(2-a)},\,a = 1- p/q$,
\be
(1+\xi^{p/q})^{ 2 q u/(p+q)} = \frac{1}{2 \pi i} \int_C d \alpha \,
\xi^{ \alpha p/q } \frac{\Gamma(-\alpha) \Gamma\left(\alpha - \frac{2
q u}{p+q}\right)}{\Gamma\left(- \frac{2q u}{p+q}\right)}.
\ee
The contour $C$ runs along the imaginary axis, to the left of
Re($\alpha$) = 0. With this we can rewrite Eq. (\ref{Idef}) in terms
of $p$ and $q$ as
\ba
{\mathcal{I}}^m & = & \int_C \frac{d \alpha}{2 \pi i}
\frac{\Gamma(-\alpha) \Gamma\left(\alpha - \frac{2 q }{q+p}
u\right)}{\Gamma\left(- \frac{2 q }{q+p} u\right)}
\frac{\Gamma\left(m-1-\frac{p}{q} \alpha\right) \Gamma\left( 1 +
\frac{(q-p)}{q+p} u + \frac{p}{q} \alpha \right)}{\Gamma\left( m +
\frac{(q-p)}{q+p} u \right)}. \label{mint}
\ea
We change variables from $\alpha$ to $\tilde{\alpha} = \alpha/q$, with
$\tilde{C}$ the contour in the new variable, and use the following
properties of the $\Gamma$-function \cite{GR,wolfram}:
\ba
\Gamma(n z) & = & n^{n z-1/2} (2 \pi)^{(1-n)/2} \prod\limits_{k =
0}^{n-1} \Gamma\left(z + \frac{k}{n} \right),\quad n \mbox{ integer},
\label{mult} \\ \Gamma(z - n) & = & \frac{(-1)^n \Gamma(z)}{(1-z)_n},
\label{inv} \\ \mbox{Res}_\alpha \!\!\!\!\!\!&
\!\!\!\!\!&\!\!\!\left\{\Gamma(\alpha - b) f(\alpha)\right\} =
\sum\limits_{n=0}^\infty \frac{(-1)^n}{n!} f(\alpha = b - n),
\label{res}
\ea 
where in (\ref{res}) the residues of the $\Gamma$-function are taken with
respect to $\alpha$.  This allows us to rewrite (\ref{mint}) as 
\ba
{\mathcal{I}}^m & = & \frac{q^{-\frac{2 q u}{q+p}} p^{m-1+
\frac{q-p}{q+p} u} (2 \pi)^{2-p-q}}{\Gamma\left(- \frac{2 q }{q+p}
u\right)\Gamma\left( m + \frac{q-p}{q+p} u \right)} \int_{\tilde{C}}
\frac{d \tilde{\alpha}}{2 \pi i} \prod\limits_{k = 0}^{q-1}
\Gamma\left(-\tilde{\alpha} + \frac{k}{q}\right) \prod\limits_{i =
0}^{p-1} \Gamma\left(- \tilde{\alpha} + \frac{m-1+i}{p}\right)
\nonumber \\ & & \quad \times\, \prod\limits_{j = 0}^{q-1}
\Gamma\left( \tilde{ \alpha} - \frac{2}{q+p} u + \frac{j}{q} \right)
\prod\limits_{h = 0}^{p-1} \Gamma\left( \tilde{\alpha} +
\frac{q-p}{(q+p) p} u + \frac{h+1}{p} \right). 
\label{mint3} 
\ea
We now close the contour to pick up the residues at $\tilde{\alpha} =
\frac{2}{p+q} u - \frac{j}{q} - n$, $n$ integer, where $j$ runs from 0
to $q-1$, and $\tilde{\alpha} = - \frac{h+ 1}{m} - \frac{q-p}{(q+p) p}
u - n$, where $h$ runs from 0 to $p-1$. We thus have $p+q$ residues,
evaluated according to (\ref{res}).  After a bit of algebra, using
Eqs. (\ref{mult}) and (\ref{inv}), as well as the definition of the
Pochhammer symbol (\ref{poch}), we can rewrite the result in terms of
generalized hypergeometric functions, which are defined as \cite{GR}
\be
\mbox{ }_p F_q ( \left\{\alpha_1, \dots, \alpha_p \right\};\left\{
\beta_1, \dots, \beta_q\right\}; z) \equiv \sum\limits_{n=0}^\infty
\frac{z^n}{n!} \frac{\prod\limits_{i=1}^p
(\alpha_i)_n}{\prod\limits_{j=1}^q (\beta_j)_n}. \label{hyperpFq}
\ee 
In order to find the poles in $u$ with respect to $a$, it is
instructive to partially rewrite the result in terms of $a$ where
possible (recall that $p = (1-a) \, q$):
\ba
{\mathcal{I}}^m \!\!\! & = \,\, {\displaystyle \sum\limits_{r=0}^{q-1}
}\!\!\! & q^{1/2-r} (2\pi)^{\frac{1-q}{2}} \prod\limits_{j=0,j\neq
r}^{q-1} \Gamma\left(\frac{j-r}{q}\right) \nonumber \\ & & \,\,
\times\, \mbox{ }_{(2-a)q} F_{(2-a)q-1}\left( \big\{ \mathcal{A}
\big\}; \big\{ \mathcal{B} \big\};(-1)^{ (2-a) q}\right) \nonumber \\
& & \,\, \times\, \,\frac{\Gamma\left(r -\frac{2}{2-a}
u\right)}{\Gamma\left(-\frac{2}{2-a} u \right)} \,
\frac{\Gamma\left((1-a)\, r + m -1 - \frac{2 (1-a)}{2-a} u\right)
\Gamma\left( 1+ u - (1-a)\, r \right)}{\Gamma\left(m + \frac{a}{2-a} u
\right)} 
\nonumber \\
 \!\!\! & \!\!\! + {\displaystyle
\sum\limits_{t=0}^{(1-a)\,q -1} } \!\!\! & q^{1/2-t} \,(1-a)^{-1/2-t}
(2\pi)^{\frac{1- {(1-a)} \,q}{2}} \prod\limits_{j=0,j\neq t}^{{
(1-a)}\,q-1} \Gamma\left(\frac{j-t}{(1-a) \,q}\right) \label{mintcoend2} \\ & &
\,\, \times\, \mbox{ }_{(2-a)q} F_{(2-a)q-1}\left( \big\{ \mathcal{A}'
\big\}; \big\{ \mathcal{B}' \big\};(-1)^{(2-a)q}\right) \nonumber \\ &
& \,\, \times\, \, \frac{\Gamma\left( t+ m + \frac{a}{2-a}
u\right)}{\Gamma\left( m + \frac{a}{2-a} u\right)} \,
\frac{\Gamma\left(\frac{1}{1-a} (t+1) + \frac{1}{1-a} \frac{a}{2-a} u
\right) \Gamma\left( - \frac{1}{1-a} (t+1) - \frac{1}{1-a} u \right)}{
\Gamma\left(-\frac{2}{2-a} u \right)} \,\,, \nonumber 
\ea

with the following arguments of the generalized hypergeometric functions
\begin{samepage}
\ba
\big\{ \mathcal{A} \big\} & \equiv &
\Big\{\alpha_1,\dots,\alpha_q,\tilde{\alpha}_1,\dots,\tilde{\alpha}_{(1-a)\,q}
\Big\} \nonumber \\ \big\{ \mathcal{B} \big\} & \equiv &
\Big\{\beta_1,\dots,\left.\beta_i\right|_{i\neq r+1},\dots,
\beta_q,\tilde{\beta}_1,\dots,\tilde{\beta}_{(1-a)\,q} \Big\} \\
\alpha_i & = & - \frac{2}{(2-a) q} u + \frac{r-1}{q} +
\frac{i}{q},\nonumber \\ \tilde{\alpha}_i & = & - \frac{2}{(2-a) q } u
+ \frac{r}{q} + \frac{m-2}{(1-a)\,q} + \frac{i}{(1-a)\,q},\nonumber \\
\beta_i & = & 1 + \frac{r + 1}{q} - \frac{i}{q} \quad i \neq r+ 1,
\nonumber \\ \tilde{\beta}_i & = & 1 - \frac{u}{(1-a)\,q} +
\frac{r}{q} - \frac{i}{(1-a)\,q}, \nonumber \\ 
\big\{ \mathcal{A'}
\big\} & \equiv &
\Big\{\alpha'_1,\dots,\alpha'_q,\tilde{\alpha}'_1,\dots,
\tilde{\alpha}'_{(1-a)\,q}\Big\}
\nonumber \\ \big\{ \mathcal{B'} \big\} & \equiv &
\Big\{\beta'_1,\dots,
\beta'_q,\tilde{\beta}'_1,\dots,\left.\tilde{\beta}'_i\right|_{i\neq
t+1},\dots,\tilde{\beta}'_{(1-a)\,q}\Big\} \\ \alpha'_i & = &
\frac{a}{2-a} \,\frac{u}{(1-a)\,q} + \frac{t+1}{(1-a)\,q} -
\frac{1}{q} + \frac{i}{q},\nonumber \\ \tilde{\alpha}'_i & = &
\frac{a}{2-a}\, \frac{u}{(1-a)\,q} + \frac{t+m - 1}{(1-a)\,q} +
\frac{i}{(1-a)\,q},\nonumber \\ \beta'_i & = & 1 + \frac{t + 1}{(1-a)\,q} 
+ \frac{u}{(1-a)\,q} + \frac{1}{q} - \frac{i}{q}, \nonumber \\
\tilde{\beta}'_i & = & 1 + \frac{t+1}{(1-a)\,q} - \frac{i}{(1-a)\,q}
\quad i \neq t + 1. \nonumber
\ea
\end{samepage}
Eq. (\ref{mintcoend2}) inserted into Eq. (\ref{abbrev}) gives the
final answer for the collinear contribution.  For $a = 0$ or
equivalently $p = q = 1$, we reproduce the collinear part of
(\ref{thrustu}), of course.

The poles in $u$ can be read off from (\ref{mintcoend2}), using the
properties of the Gamma- and hypergeometric functions involved
\cite{GR,wolfram}. The $\Gamma$-function $\Gamma(z)$ has simple poles at
$z = -n,\, n$ integer, with residues $(-1)^n/n!$, see Eq. (\ref{res}),
the hypergeometric $\mbox{ }_p F_q$ (\ref{hyperpFq}) has simple poles
only in $\beta_j$ at $\beta_j = -n,\,n$ integer, with residues
\be
\frac{(-1)^n}{n!} \mbox{ }_p \tilde{F}_q
\left(\left\{\alpha_i\right\};\left\{\beta_1,\dots,\beta_{j-1},-n,
\beta_{j+1},\dots,\beta_q\right\};z\right),
\ee
where $\mbox{ }_p \tilde{F}_q$ denotes the regularized hypergeometric function 
\be
\mbox{ }_p \tilde{F}_q ( \left\{\alpha_1, \dots, \alpha_p
\right\};\left\{ \beta_1, \dots, \beta_q\right\}; z) \equiv
\sum\limits_{n=0}^\infty \frac{z^n}{n!} \frac{\prod\limits_{i=1}^p
(\alpha_i)_n}{\prod\limits_{j=1}^q \Gamma(n + \beta_j)}.
\ee
This means, that despite its appearance, (\ref{mintcoend2}) has poles
in $u$ only in the $\Gamma$-functions, and in the $\tilde{\beta}_i$ and
$\beta'_i$. Note that any pole in $u$ at $u = \rho$ is accompanied by
a factor of $\nu^{\frac{2 \rho}{2-a}}$, resulting in general in a
contribution of order $\mathcal{O} \left( \nu^{\frac{1}{2-a}}
(\LQCD/Q)\right)^{2 \rho}$.

\end{appendix}
\vspace*{2mm}

\subsection*{Acknowledgements}
\vspace*{2mm}

\noindent We thank Mauro Anselmino, Einan Gardi, and George Sterman for helpful
comments. L.M. thanks the CERN PH Department (Theory
Unit) for hospitality during part of this work.

\end{document}